\documentclass[12pt]{iopart}
\usepackage{iopams}  
\usepackage{bm}
\usepackage{bbm}
\usepackage{graphics}

\begin{document}

\title[Photon propagation through linearly active dimers]{Photon propagation through linearly active dimers}

\author{Jos\'e Delfino Huerta Morales$^{1}$ and B. M. Rodr\'iguez-Lara$^{2,1}$}

\address{$^{1}$ \quad Instituto Nacional de Astrof\'{\i}sica, \'Optica y Electr\'onica, Calle Luis Enrique Erro No. 1, Sta. Ma. Tonantzintla, Pue. CP 72840, M\'exico; jd\_huerta@inaoep.mx\\
	$^{2}$ \quad Photonics and Mathematical Optics Group, Tecnol\'ogico de Monterrey, Monterrey 64849, Mexico.}
\ead{bmlara@itesm.mx}
\vspace{10pt}
%\begin{indented}
%\item[]February 2014
%\end{indented}

\begin{abstract}
We provide an analytic propagator for non-Hermitian dimers showing linear gain or losses in the quantum regime. In particular, we focus on experimentally feasible realizations of the $\mathcal{PT}$-symmetric dimer and provide their mean photon number and second order two-point correlation. We study the propagation of vacuum, single photon spatially-separable, and two-photon spatially-entangled states. We show that each configuration produces a particular signature that might signal their possible uses as photon switches, semi-classical intensity-tunable sources, or spatially entangled sources to mention a few possible applications.\end{abstract}

%\maketitle

\section{Introduction}

Propagation through classical $\mathcal{PT}$-symmetric optical systems has been extensively studied; c.f. Ref.~\cite{HuertaMorales2016p83} and references therein from the initial description of linear loses in directional couplers with a more complex non-Hermitian symmetry \cite{Somekh1973p46}, the quantum-like description of classical planar waveguides \cite{Ruschhaupt2005p171}, to all the contemporaneous work derived from the seminal introduction of optical $\mathcal{PT}$-symmetric structures \cite{ElGanainy2007p2632}.
It is well known that the propagation of electromagnetic field through a linearly active two-waveguide coupler can be described by the classical $\mathcal{PT}$-symmetric dimer, 
\begin{eqnarray}
-i\frac{d}{dz} \left( \begin{array}{c} \mathcal{E}_{1}(z) \\ \mathcal{E}_{2}(z) \end{array}\right) = \left( \begin{array}{cc} i \gamma & g \\ g & -i \gamma  \end{array} \right) \left( \begin{array}{c} \mathcal{E}_{1}(z) \\ \mathcal{E}_{2}(z) \end{array}\right),
\end{eqnarray}
where the effective evanescent coupling between the two single waveguide field modes is given by the real parameter $g$ and the effective gain and loss by the real positive parameter $\gamma$.
Such a system can be realized experimentally by balanced gain and loss in the waveguides, but it is also possible to realize it with gain-gain, loss-loss, passive-gain \cite{Ruter2010p192}, and passive-loss \cite{Guo2009p093902,Ornigotti2014p065501} waveguide, microcavity rings \cite{Peng2014p394,Hodaei2014p975}, and electric circtuits \cite{Schindler2012p444029} setups.

\begin{figure}[h]
	\centering
	\includegraphics{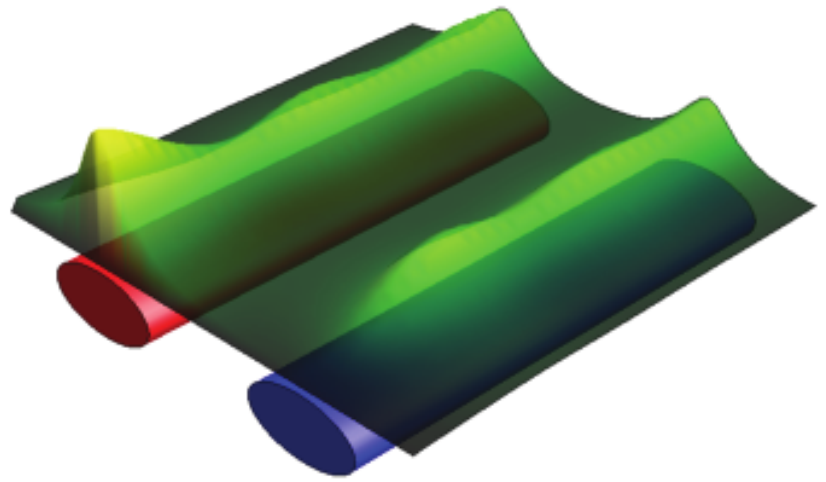}
	\caption{Schematic showing the renormalized light intensity arising from the spontaneous generation of photons through two coupled waveguides in the $\mathcal{PT}$-symmetry regime with balanced gain-loss~configuration.} \label{fig:Fig1}
\end{figure}

In the quantum regime \cite{Politi2008p646,Bromberg2009p253904,Peruzzo2010p1500,Joglekar2013p30001},  the importance of adequately modeling media with linear gain or loss has been brought forward recently \cite{Agarwal2012p031802,Grafe2013p033008}. 
In this regime, linear media induces quantum fluctuations, such that the ideal $\mathcal{PT}$-symmetric optical dimer dynamics is effectively described by quantum Langevin equations \cite{Agarwal2012p031802},  
\begin{eqnarray}
-i \frac{d}{dz} \left( \begin{array}{c} \hat{a}_{1}(z) \\ \hat{a}_{2}(z) \end{array}\right) = \left( \begin{array}{cc} i \gamma & g \\ g & -i \gamma  \end{array} \right) \left( \begin{array}{c} \hat{a}_{1}(z) \\ \hat{a}_{2}(z) \end{array}\right) + \left( \begin{array}{c} \hat{f}_{1}(z) \\ \hat{f}_{2}(z) \end{array}\right),
\end{eqnarray}
in terms of the effective balanced gain and loss parameter, $\gamma$, the effective mode coupling, $g$, between the two modes of the optical resonators, described by the annihilation operators $\hat{a}_{1}(z)$ and $\hat{a}_{2}(z)$, and the delta-correlated Langevin forces introduced by the gain and loss media, 
\begin{eqnarray}
\langle \hat{f}_{1}^{\dagger}(z) \hat{f}_{1}(\zeta) \rangle = 2 \gamma \delta(z-\zeta), \nonumber \\
\langle \hat{f}_{2}(z) \hat{f}_{2}^{\dagger}(\zeta) \rangle = 2 \gamma \delta(z-\zeta),
\end{eqnarray} 
in that order.
These Gaussian fluctuations modify the well-known dynamics produced by the classical optical $\mathcal{PT}$-symmetric dimer and generate second order correlations in the ideal quantum optical $\mathcal{PT}$-symmetric dimer \cite{Agarwal2012p031802}, Fig. \ref{fig:Fig1}. Here, we want to discuss the different types of correlations that might arise from feasible experimental realizations of the quantum $\mathcal{PT}$-symmetric dimer beyond the balanced gain-loss setup.
In the following section, we will introduce the quantum model for a generalized linear active dimer, a non-Hermitian quantum optical dimer, and provide its propagation solution. Then, we will discuss the dynamics of spontaneous photon generation as well as photon bunching and anti-bunching in the different configurations. We will study single-photon propagation with spatially separable and two-photon propagation with entangled states through the mean photon number and second order two-point correlations. Finally, we will close with a brief conclusion and discuss how mean photon propagation and second order spatial correlations might provide insight to their application in photon switching or as semi(non)-classical light sources.

%%%%%%%%%%%%%%%%%%%%%%%%%%%%%%%%%%%%%%%%%%%%%%%%%%%%%%%%%%%%%%%%%%%%%%%%%%%%%%%%%%%%%%%%%%%%%%%%%%	
\section{Quantum model and configurations}
%%%%%%%%%%%%%%%%%%%%%%%%%%%%%%%%%%%%%%%%%%%%%%%%%%%%%%%%%%%%%%%%%%%%%%%%%%%%%%%%%%%%%%%%%%%%%%%%%%
In the laboratory, we can think about a more general realization of the effective quantum $\mathcal{PT}$-symmetric dimer provided by the quantum non-Hermitian dimer,
\begin{eqnarray}
	-i\frac{d}{dz} \left( \begin{array}{c} \hat{a}_{1}(z) \\ \hat{a}_{2}(z) \end{array}\right) = \left( \begin{array}{cc} n_{1} & g \\ g & n_{2}  \end{array} \right) \left( \begin{array}{c} \hat{a}_{1}(z) \\ \hat{a}_{2}(z) \end{array}\right) + \left( \begin{array}{c} \hat{f}_{1}(z) \\ \hat{f}_{2}(z) \end{array}\right).
\end{eqnarray}
Again, the field annihilation operators in the first and second waveguides are given by the operators $\hat{a}_{1}(z)$ and $\hat{a}_{2}(z)$, in that order.
The effective refractive indices of the optical waveguides are given by the complex numbers $n_{1}$ and $n_{2}$, the effective coupling between the optical modes is given by the real positive parameter $g$. Finally, the Gaussian fluctuations, due to the active linear media, are described by the Langevin forces $\hat{f}_{1}(z)$ and $\hat{f}_{2}(z)$, such that the only nonzero mean values involving them are their second order correlations \cite{Scully2001}, 
\begin{eqnarray}
	\langle \hat{f}_{j}^{\dagger}(z) \hat{f}_{j}(z') \rangle &=& 2 \Im({n}_{j}) \delta( z - z'), \quad \mathrm{for~gain~media}, \nonumber  \label{1124}\\
	\langle \hat{f}_{j}(z) \hat{f}_{j}^{\dagger}(z') \rangle &=& 2 \Im({n}_{j}) \delta( z - z'), \quad \mathrm{for~loss~media}. \label{1125}
\end{eqnarray}

It is straightforward to write a formal propagator for this quantum optical system,
\begin{eqnarray}
	\left( \begin{array}{c} \hat{a}_{1}(\zeta) \\ \hat{a}_{2}(\zeta) \end{array}\right) =  e^{i n_{0} \zeta} \left( \begin{array}{c} \hat{b}_{1}(\zeta) \\ \hat{b}_{2}(\zeta) \end{array}\right), 
\end{eqnarray}
where we have factorized an average refractive index, $n_{0} = (n_{1} + n_{2})/(2 g)$, that introduces a common phase, $\Re(n_{0})$, and a scaling factor due to gain or loss, $\Im(n_{0})$. We have also scaled the propagation variable by the effective coupling of the dimer, $\zeta = g z$.
Note that the average refractive index cannot be zero for experimental realizations. It becomes a pure real number, a phase factor, for passive materials, and a pure imaginary number, a scaling factor, for identical media with balanced gain and loss.
The~second term in the propagator,
\begin{eqnarray} \label{305_7}
	\left( \begin{array}{c} \hat{b}_{1}(\zeta) \\ \hat{b}_{2}(\zeta) \end{array}\right) =  e^{i \hat{H} \zeta} \left( \begin{array}{c} \hat{b}_{1}(0) \\ \hat{b}_{2}(0) \end{array}\right) + \int_{0}^{\zeta} e^{ i \hat{H} (\zeta - t) } e^{-i n_{0} t} \left( \begin{array}{c} \hat{f}_{1}(t) \\ \hat{f}_{2}(t) \end{array}\right) dt, 
\end{eqnarray}
provides us with the effective dynamics of the system.
Here, we need use the scaling property of Dirac delta, $\delta( \zeta) = \delta(z)/\vert g \vert $, and have defined a complex effective refractive index $n =  (n_{1} - n_{2})/(2 g)$ that provides us with an auxiliary effective non-Hermitian matrix,
\begin{eqnarray}
	\hat{H} = \left( \begin{array}{cc} n & 1 \\ 1 & -n \end{array} \right).
\end{eqnarray}
Note that the case of waveguides with identical real part of their effective refractive indices, $\Re(n_{1}) = \Re(n_{2})$, yields a purely imaginary effective refractive index that we can rename as $n =  i \gamma$ in order to recover the standard quantum $\mathcal{PT}$-symmetric dimer \cite{Agarwal2012p031802}. 
Again, let us stress that the dynamics introduced by a more realistic model of linearly active media, where configurations beyond balanced gain-loss are easily obtained, will include a phase and scaling factor proportional to the average refractive index that are not taken into consideration in the ideal $\mathcal{PT}$-symmetric dimer configuration.

The coupling matrix exponential can be easily calculated following an approach similar to that used in the classical $\mathcal{PT}$-symmetric dimer \cite{HuertaMorales2016p83},
\begin{equation}
	\begin{array}{lll}
		\hat{U}(\zeta) &=& e^{i \hat{H} \zeta}, \\
		&=& \hat{1} \cos \left( \Omega \zeta \right) +  \hat{H}~ \zeta~ \mathrm{sinc} \left( \Omega \zeta \right), \qquad \Omega \in \mathbb{C}.
	\end{array}
\end{equation}
Here, the symbol $\hat{1}$ stands for the two by two identity matrix, we have used the cardinal sine function $\mathrm{sinc}(x) = \sin(x) /x$,  and the complex dispersion relation is given by the following expression,
\begin{eqnarray}
	\Omega = \sqrt{ 1 + n^2}.
\end{eqnarray}
Note that this analytic propagator can describe any non-Hermitian dimer and, in the special case of  purely imaginary auxiliary refractive index, $n = i \gamma$, we recover the propagator for the standard classical $\mathcal{PT}$-symmetric dimer \cite{RodriguezLara2015p5682},
\begin{eqnarray}
	\hat{U}(\zeta) =  \left\{ \begin{array}{ll} 
		\hat{1} \cos \left( \Omega \zeta \right) +  i~\hat{H}~ \zeta~  \mathrm{sinc} \left( \Omega \zeta \right) & \gamma < 1, \\
		\hat{1} +  i ~\zeta~ \hat{H}, & \gamma = 1, \\
		\hat{1} \cosh \left( \vert \Omega \vert \zeta \right) +  i ~ \hat{H}~ \zeta~  \mathrm{sinhc} \left( \vert \Omega \vert \zeta \right), & \gamma > 1.
	\end{array}  \right.
\end{eqnarray}

Experimentally, the effective $\mathcal{PT}$-symmetric dimer can be reached via different configurations: balanced gain-loss dimer, $n_{1} = n_{R} - i n_{I}$ and $n_{2} = n_{R} + i n_{I}$, in that order, such that \mbox{$n_{0} = n_{R}/ g$} and \mbox{$\gamma = -n_{I}/g$}; gain-gain dimer, $n_{1} = n_{R} - i n_{I,1}$ and $n_{2} = n_{R} - i n_{I,2}$ such that \mbox{$n_{0} = \left[2 n_{R} -i \left( n_{I,1} + n_{I,2}\right) \right] / (2 g)$} and $\gamma = - (n_{I,1} - n_{I,2})/(2g)$; passive-gain dimer, $n_{1} = n_{R}$ and $n_{2} = n_{R} - i n_{I}$, in that order, such that $n_{0} = (2 n_{R} - i n_{I})/ (2 g)$ and $\gamma = n_{I}/(2g)$; passive-loss dimer, $n_{1} = n_{R}$ and $n_{2} = n_{R} + i n_{I}$, in that order, such that $n_{0} = (2 n_{R} + i n_{I})/ (2 g)$ and $\gamma =- n_{I}/(2g)$; and loss-loss dimer, $n_{1} = n_{R} + i n_{I,1}$ and $n_{2} = n_{R} + i n_{I,2}$ such that $n_{0} = \left[2 n_{R} +i (n_{I,1} + n_{I,2}) \right]/ (2 g)$ and $\gamma = (n_{I,1} - n_{I,2})/(2g)$; for all of these configurations, we have assumed $n_{R}$, $n_{I}$, $n_{I,j}>0$. 
The~formal solution presented in this section allows us to explore the propagation properties of all of these experimentally feasible configurations.
Table \ref{tab:Tab1} shows the values of these parameters for the different configurations delivering an effective $\mathcal{PT}$-symmetric dimer; and we have added the imaginary part of the effective bias refractive index, $\beta = \left( n_{I,1} + n_{I,2} \right)/ (2g)=~ - \Im(n_{0}) $, as it will be useful in the following~sections. 

	\begin{table}[h]
		\caption{A summary of the parameters involved in the different feasible experimental realizations of the $\mathcal{PT}$-symmetric dimer .} \label{tab:Tab1}
		%\centering
		%\tablesize{\footnotesize} %% You can specify the fontsize here, e.g.  \tablesize{\footnotesize}. If commented out \small will be used.
		\begin{tabular}{lllcccc}
			\hline
		\textbf{Realization}	& $\mathbf{n_{1}}$	& $\mathbf{n_{2}}$ & $\mathbf{n}$ & $\mathbf{n_{0}}$  & $\bm{\gamma}$  & $\bm{\beta}$ \\
			\hline
			\textbf{Gain-loss}	&  $n_{R} - i n_{I}$ &  $n_{R} + i n_{I}$ & $-i\frac{n_{I}}{g}$ &  $\frac{n_{R}}{g}$ & $-\frac{n_{I}}{g}$ & 0\\
			\textbf{Gain-gain}	&  $n_{R} - i n_{I,1}$ &  $n_{R} - i n_{I,2}$ & $i\frac{-n_{I,1}+n_{I,2}}{2g}$ &  $\frac{n_{R}}{g}-i\frac{n_{I,1}+n_{I,2}}{2g}$ & $\frac{-n_{I,1}+n_{I,2}}{2g}$ & $\frac{n_{I,1}+n_{I,2}}{2g}$\\
			\textbf{Gain-passive}	&  $n_{R}- i n_{I}$ &  $n_{R} $ & $-i\frac{n_{I}}{2g}$ &  $\frac{n_{R}}{g}+i\frac{n_{I}}{2g}$ & $-\frac{n_{I}}{2g}$ & $\frac{n_{I}}{2g}$						\\
			\textbf{Passive-loss}  &  $n_{R}$ &  $n_{R} + i n_{I}$ & $-i\frac{n_{I}}{2g}$ &  $\frac{n_{R}}{g}+i\frac{n_{I}}{2g}$ & $-\frac{n_{I}}{2g}$ & $-\frac{n_{I}}{2g}$  \\
			\textbf{Loss-loss} & $n_{R}+ i n_{I,1}$ &  $n_{R} + i n_{I,2}$ & $i \frac{n_{I,1}-n_{I,2}}{2g}$ &  $\frac{n_{R}}{g} + i \frac{n_{I,1}+n_{I,2}}{2g}$ & $\frac{n_{I,1}-n_{I,2}}{2g}$  & $\frac{-n_{I,1}-n_{I,2}}{2g}$					\\	                                      		
			\hline
		\end{tabular}
		\label{tab:Tab1}
	\end{table}

%%%%%%%%%%%%%%%%%%%%%%%%%%%%%%%%%%%%%%%%%%%%%%%%%%%%%%%%%%%%%%%%%%%%%%%%%%%%%%%%%%%%%%%%%%%%%%%%%%	
\section{Spontaneous generation of photons} \label{sec:SpontaneousGeneration}
%%%%%%%%%%%%%%%%%%%%%%%%%%%%%%%%%%%%%%%%%%%%%%%%%%%%%%%%%%%%%%%%%%%%%%%%%%%%%%%%%%%%%%%%%%%%%%%%%%

The first signature that differentiates a quantum from a classical $\mathcal{PT}$-symmetric dimer is the spontaneous generation of photons due to the gain medium in presence of vacuum input fields,
\begin{equation}
\begin{array}{lll}
n^{(00)}_{j}(\zeta) &=& \langle 0,0 \vert \hat{a}_{j}^{\dagger}(\zeta) \hat{a}_{j}(\zeta) \vert 0,0 \rangle,  \\
&=& e^{2 \beta \zeta}  \langle 0,0 \vert \hat{b}_{j}^{\dagger}(\zeta) \hat{b}_{j}(\zeta) \vert 0,0 \rangle.
\end{array}
\end{equation}
In other words, vacuum fluctuations are enough to make the linear active media spontaneously generate photons \cite{Scully2001}; something that is lacking in the classical model.
The signatures available through the spontaneous generation of photos in diverse configurations are provided in the following for the standard balanced \textbf{gain-loss} dimer,
\begin{equation}
\begin{array}{lll}
n^{(00)}_{1}(\zeta) &=&- 2 \gamma {\int_{0}^{\zeta}{\left| \hat{U}_{11}\left( t \right) \right|}^2} dt,   \\
n^{(00)}_{2}(\zeta) &=& -2\gamma {\int_{0}^{\zeta}{\left| \hat{U}_{21}\left( t \right) \right|}^2}dt, \label{703_14}
\end{array}
\end{equation}
where the (i,j)-th component of the two by two propagation matrix has been written as $\hat{U}_{ij}(\zeta)$, and it is important to note that these integrals can be solved analytically but yield expressions too long to write here.
The \textbf{gain-gain} dimer yields the following expressions,
\begin{equation}
\begin{array}{lll}
n^{(00)}_{1}(\zeta) &=& 2 \left( \beta - \gamma \right) {\int_{0}^{\zeta}{\left| \hat{U}_{11}\left( t \right) \right|}^2} e^{2 \beta t} dt ~ + ~  2 \left( \beta + \gamma \right) {\int_{0}^{\zeta}{\left| \hat{U}_{12}\left( t \right) \right|}^2} e^{2 \beta t} dt    ,\\
n^{(00)}_{2}(\zeta) &=&  2 \left( \beta - \gamma \right) {\int_{0}^{\zeta}{\left| \hat{U}_{21}\left( t \right) \right|}^2} e^{2 \beta t} dt ~+~ 2 \left( \beta + \gamma \right) {\int_{0}^{\zeta}{\left| \hat{U}_{22}\left( t \right) \right|}^2} e^{2 \beta t} dt . \label{eq:SGGG}
\end{array}
\end{equation}
For the \textbf{gain-passive} dimer, we can write the spontaneous generation as:
\begin{equation}
\begin{array}{lll}
n^{(00)}_{1}(\zeta) &=& -4\gamma {\int_{0}^{\zeta}{\left| \hat{U}_{11}\left( t \right) \right|}^2}  e^{-2 \gamma t} dt, \\
n^{(00)}_{2}(\zeta) &=& -4\gamma {\int_{0}^{\zeta}{\left| \hat{U}_{21}\left( t \right) \right|}^2} e^{-2 \gamma t} dt, \label{eq:SGGP} 
\end{array}
\end{equation}
and, obviously, there is not spontaneous generation in the \textbf{passive-loss} and \textbf{loss-loss} dimer, 
\begin{eqnarray}
n^{(00)}_{1}(\zeta) = n^{(00)}_{2}(\zeta) = 0. \label{eq:SGPLLL}
\end{eqnarray}
While the expressions for the spontaneous generation are complicated, it is straightforward to realize from the analytic expressions that they will present different signatures through propagation in the dimer.

\begin{figure}[h]
	\centering
	\includegraphics{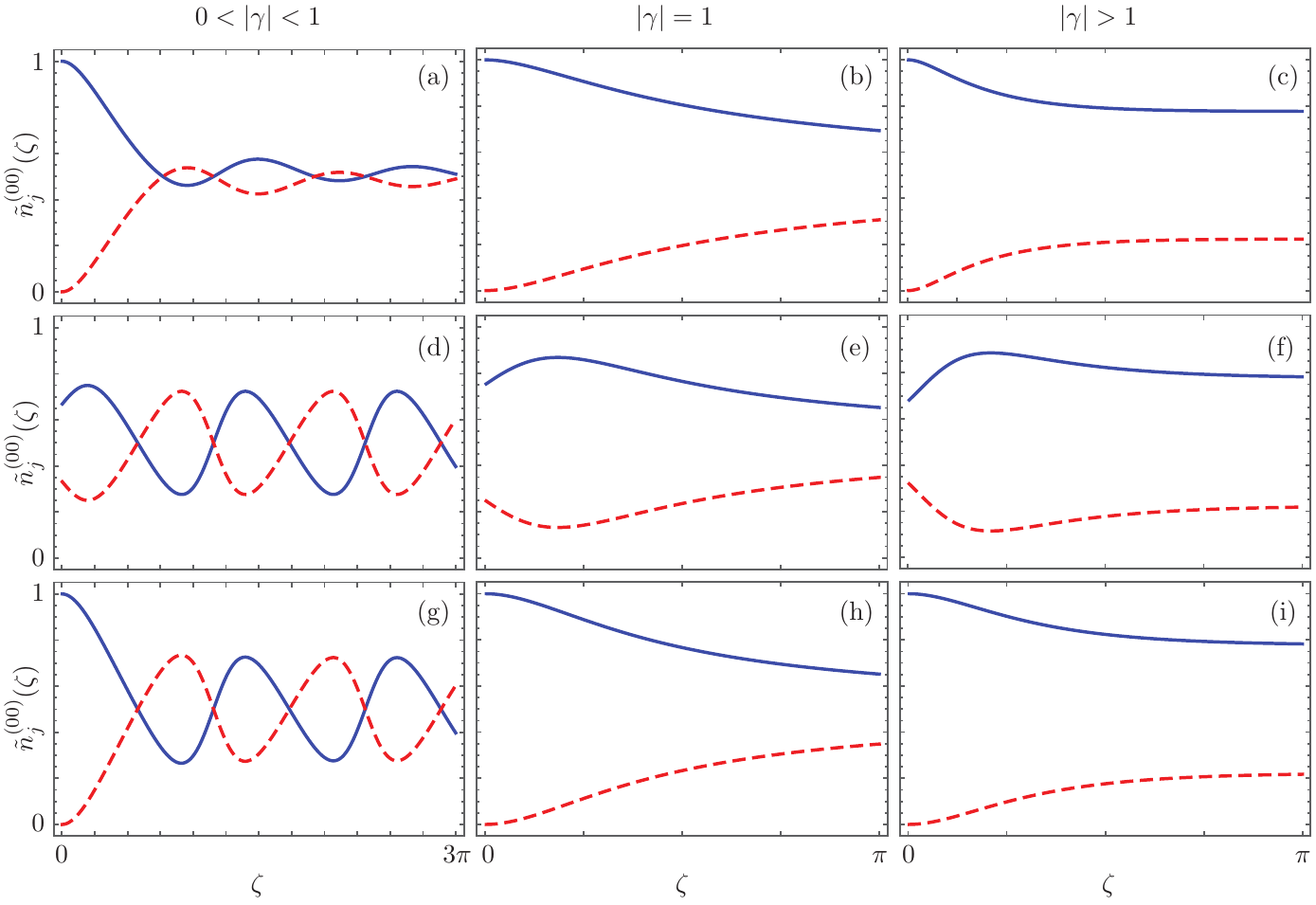}
	\caption{Instantaneously renormalized spontaneous generation, $\tilde{n}_{j}^{(00)}(\zeta)$, along different realizations of the effective $\mathcal{PT}$-symmetric dimer. The first row, (a)-(c), shows balanced gain-loss, second row, (d)-(f), shows gain-gain, and third row, (g)-(i), shows gain-passive configurations in the $\mathcal{PT}$-symmetric regime, the first column with $\vert \gamma \vert =0.5$, the Kato point with $\vert \gamma \vert = 1$, the second column, and broken symmetry regime, and the third column with $\vert \gamma \vert =1.2 $. Values for the first and second waveguides are shown with a solid blue and a dashed red lines, in that order. Note the oscillatory behavior of the spontaneous generation inside and its asymptotic behavior outside the $\mathcal{PT}$-symmetric regime.}%there is explanation for subfigure. Please add the explanation of (a)-(i) in the caption.
	\label{fig:Fig2} 
\end{figure}

The signatures from the different configurations can be seen in Fig. \ref{fig:Fig2}, where we show an instantaneously renormalized spontaneous generation of photons,
\begin{eqnarray}
\tilde{n}_{j}^{(00)}(\zeta) = \frac{n_{j}^{(00)}(\zeta)}{n_{1}^{(00)}(\zeta)+n_{2}^{(00)}(\zeta)}, \qquad \zeta > 0, 
\end{eqnarray}
and avoid the position $\zeta=0$ due to the divergence induced by the initial vacuum field.
The rows in Fig.~\ref{fig:Fig2} present the spontaneous emission in the balanced gain-loss, gain-gain, gain-passive configurations, from top to bottom, and the columns show results in the $\mathcal{PT}$-symmetric, Kato exceptional point, and broken symmetry regimes, from left to right. 
The spontaneous emission in the ideal $\mathcal{PT}$-symmetric is shown in Fig. \ref{fig:Fig2}(a)--\ref{fig:Fig2}(c). 
Note that the oscillations in the $\mathcal{PT}$-symmetric regime appear earlier in the propagation for the gain-gain, Fig. \ref{fig:Fig2}(d), and gain-passive configurations, Fig.~\ref{fig:Fig2}(g).
In the Kato exceptional point, the spontaneous emission is equivalent in the ideal dimer, Fig.~\ref{fig:Fig2}(b), and the gain-passive configurations, Fig.~\ref{fig:Fig2}(h), and follows a slightly different initial distribution in the gain-gain case Fig. \ref{fig:Fig2}(e). 
Finally, something similar happens in the broken symmetry regime, the distinction between spontaneous emission in the waveguides  for the ideal, Fig. \ref{fig:Fig2}(c), and gain-passive configurations, Fig.~\ref{fig:Fig2}(i), is almost null and follows a slightly different initial distribution for the gain-gain case, Fig.~\ref{fig:Fig2}(f).
Note that the spontaneous generation is asymptotically equal in both waveguides when the dimer is at the Kato exceptional point for any configuration with linear gain, Fig.~\ref{fig:Fig2}(b), \ref{fig:Fig2}(e), \ref{fig:Fig2}(h). 
In addition, in the broken symmetry regime, the asymptotic value for the renormalized spontaneous generation is the same for the different configurations including linear gain, Fig. \ref{fig:Fig2}(c), \ref{fig:Fig2}(f), \ref{fig:Fig2}(i), and converges to the same value than the classical $\mathcal{PT}$-symmetric dimer \cite{HuertaMorales2016p83,RodriguezLara2015p5682},
\begin{equation}
\begin{array}{lll}
\lim_{\zeta \rightarrow \infty} \tilde{n}_{1}^{(00)} (\zeta) &=& \frac{1}{2 \gamma} \left( \gamma + \sqrt{ \gamma^2 - 1} \right)^{-1}, \\
\lim_{\zeta \rightarrow \infty} \tilde{n}_{2}^{(00)} (\zeta) &=& \frac{1}{2 \gamma} \left( \gamma + \sqrt{ \gamma^2 - 1} \right).
\end{array}
\end{equation}

%%%%%%%%%%%%%%%%%%%%%%%%%%%%%%%%%%%%%%%%%%
\section{Photon bunching in spontaneous generation}
%%%%%%%%%%%%%%%%%%%%%%%%%%%%%%%%%%%%%%%%%%

We can also look at the signatures provided by the probability to detect simultaneously one photon at each of the waveguides output in the different configurations. This can be written in the following form for the spontaneous generation of photons,
\begin{eqnarray}
\left\langle 0,0 \vert {{\hat{a}_{1}}^{\dagger }}\left( \zeta \right){{\hat{a}_{2}}^{\dagger }}\left( \zeta \right) \hat{a}_{1}\left( \zeta \right)\hat{a}_{2}\left( \zeta \right) \vert 0, 0 \right\rangle &=& \left\langle 0,0 \vert {{\hat{a}_{1}}^{\dagger }}\left( \zeta \right) \hat{a}_{1}\left( \zeta \right) \vert 0, 0 \right\rangle \left\langle 0,0 \vert {{\hat{a}_{2}}^{\dagger }}\left( \zeta \right) \hat{a}_{2}\left( \zeta \right) \vert 0, 0 \right\rangle \nonumber \\ 
& +& \left\langle 0,0 \vert {{\hat{a}_{1}}^{\dagger }}\left( \zeta \right) \hat{a}_{2}\left( \zeta \right) \vert 0, 0 \right\rangle \left\langle 0,0 \vert {{\hat{a}_{2}}^{\dagger }}\left( \zeta \right) \hat{a}_{1}\left( \zeta \right) \vert 0, 0 \right\rangle \nonumber \\ 
&=& n^{(00)}_{1}(\zeta) n^{(00)}_{2}(\zeta) + \left | n^{(00)}_{12}(\zeta) \right|^2,
\end{eqnarray} 
where he have used the Gaussian nature of Langevin forces  \cite{Agarwal2012p031802}  to simplify this probability using the expressions derived in the last section and defining the following first order two-point correlation,
\begin{equation}
\begin{array}{lll}
n^{(00)}_{12}(\zeta) &=& \langle 0,0 \vert \hat{a}_{1}^{\dagger}(\zeta) \hat{a}_{2}(\zeta) \vert 0,0 \rangle,  \\
&=& e^{2 \beta \zeta}  \langle 0,0 \vert \hat{b}_{1}^{\dagger}(\zeta) \hat{b}_{2}(\zeta) \vert 0,0 \rangle.
\end{array}
\end{equation}
Note that the detection probability will always be positive independently of the form taken by the first order two-point correlation for the different configurations: \textbf{balanced gain-loss} dimer,
\begin{eqnarray}  \label{703_23}
n^{(00)}_{12}(\zeta) &=& -2 \gamma \int_{0}^{\zeta} \hat{U}_{11}^{\ast} \left( t \right) \hat{U}_{21} \left( t \right) dt,
\end{eqnarray}
\textbf{gain-gain} dimer,
\begin{eqnarray} \label{eq:1003_24}
n^{(00)}_{12}(\zeta) &=& 2 \left( \beta - \gamma \right)  \int_{0}^{\zeta}{\hat{U}_{11}^{\ast} \left( t \right) \hat{U}_{21} \left( t \right)}  e^{2 \beta t} dt \nonumber \\ && + 2 \left( \beta + \gamma \right)  \int_{0}^{\zeta}{\hat{U}_{12}^{\ast} \left( t \right) \hat{U}_{22} \left( t \right)}  e^{2 \beta t} dt,
\end{eqnarray}
\textbf{gain-passive} dimer,
\begin{eqnarray} 
n^{(00)}_{12}(\zeta) &=& -4 \gamma \int_{0}^{\zeta}{\hat{U}_{11}^{\ast} \left( t \right) \hat{U}_{21} \left( t \right)}  e^{-2 \gamma t} dt, \label{eq:12GP}
\end{eqnarray}
and, obviously, for the \textbf{passive-loss} and \textbf{loss-loss} dimer,
\begin{eqnarray}
n^{(00)}_{12}(\zeta) = 0.
\end{eqnarray}
Again, we have to be careful to consider the appropriate parameters $\gamma$ and $\beta$ for each configuration summarized in Table \ref{tab:Tab1}. 

\begin{figure}[h]
	\centering
	\includegraphics{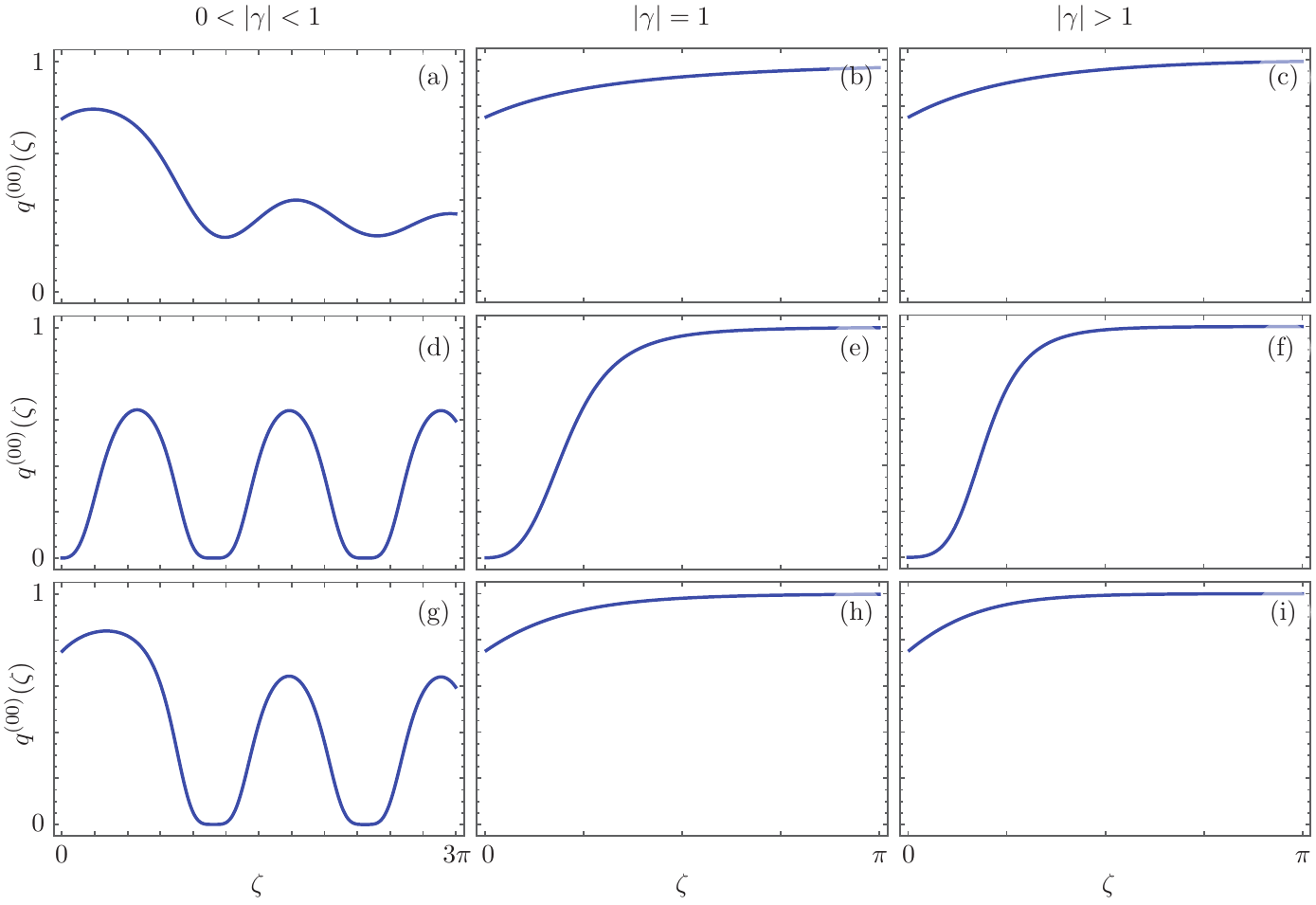}
	\caption{Photon bunching shown in terms of the $q^{(00)}(\zeta)$ parameter for different realizations of the effective $\mathcal{PT}$-symmetric dimer. The first row, (a)-(c), shows balanced gain-loss, second row, (d)-(f), shows gain-gain, and third row, (g)-(i), shows gain-passive configurations in the $\mathcal{PT}$-symmetric regime, the first column with $\vert \gamma \vert =0.5$, the Kato point with $\vert \gamma \vert = 1$, the second column, and broken symmetry regime, and the third column with $\vert \gamma \vert =1.2 $.}
	\label{fig:Fig3}
\end{figure}

In order to visualize the information, we can use a quantity similar to Mandel Q-parameter \cite{Mandel1979p205},
\begin{equation}
\begin{array}{lll}
q^{(00)}(\zeta) &=& g_{2}^{(00)}(\zeta)-1, \\
&=&  \frac{\vert n_{12}^{(00)}(\zeta) \vert^2}{n^{(00)}_{1}(\zeta) n^{(00)}_{2}(\zeta)}, \qquad \zeta > 0,
\end{array}
\end{equation}
given in terms of the second order two-point correlation function, 
\begin{eqnarray}
g_{2}^{(00)}(\zeta) = 1 +  \frac{\vert n_{12}^{(00)}(\zeta) \vert^2}{n^{(00)}_{1}(\zeta) n^{(00)}_{2}(\zeta)},  \qquad \zeta > 0,
\end{eqnarray}
that provides us with the probability of simultaneously detecting a photon in each waveguide output. 
The values of this two-point parameter are always positive, $q^{(00)}(\zeta)~\ge~0$, thus, the different configurations only show photon bunching.
Figure \ref{fig:Fig3} shows the photon bunching signatures obtained with the different dimer configurations discussed before. 
The ordering is the same than in Fig. \ref{fig:Fig2}, rows show the balanced gain-loss, gain-gain, and gain-passive dimer configurations, in that order, and columns show the symmetric, exceptional and broken symmetry regimes from left to right.
We can immediately see that the signatures in the $\mathcal{PT}$-symmetric regime, Fig.~\ref{fig:Fig3}(a), \ref{fig:Fig3}(d), \ref{fig:Fig3}(g), present an oscillatory behavior, while those in the Kato exceptional point, Fig.~ \ref{fig:Fig3}(b), \ref{fig:Fig3}(e), \ref{fig:Fig3}(h), and the broken $\mathcal{PT}$-symmetry regimes, and Fig.~ \ref{fig:Fig3}(c), \ref{fig:Fig3}(f), \ref{fig:Fig3}(i), saturate to the unit. 
The dimer configuration also influences the photon bunching signature, the balanced gain-loss dimer presents oscillations that do not approach zero, Fig. \ref{fig:Fig3}(a), while the gain-gain and the gain-passive dimers present oscillations that reach zero, Fig.~\ref{fig:Fig3}(d), \ref{fig:Fig3}(g). 
In addition, the balanced gain-loss dimer saturates in a manner similar to that of the gain-passive dimer starting from a nonzero value, Fig.~\ref{fig:Fig3}(b), \ref{fig:Fig3}(h) as well as Fig.~\ref{fig:Fig3}(c), \ref{fig:Fig3}(i), while the gain-gain dimer starts from zero and saturates faster, Fig. \ref{fig:Fig3}(e), \ref{fig:Fig3}(f).

%%%%%%%%%%%%%%%%%%%%%%%%%%%%%%%%%%%%%%%%%%
\section{Photon propagation}
%%%%%%%%%%%%%%%%%%%%%%%%%%%%%%%%%%%%%%%%%%

In general, we can study the propagation of any given initial photon state through our dimer,
\begin{equation}
\begin{array}{lll}
n^{(\psi_{0})}_{j}(\zeta) &=& \langle \psi_{0} \vert \hat{a}_{j}^{\dagger}(\zeta) \hat{a}_{j}(\zeta) \vert \psi_{0} \rangle, \\
&=& e^{2 \beta \zeta}  \langle \psi_{0} \vert \hat{b}_{j}^{\dagger}(\zeta) \hat{b}_{j}(\zeta) \vert \psi_{0} \rangle, \label{eq:MeanPhotonNumber}
\end{array}
\end{equation}
and realize that the mean photon number in each waveguide,
\begin{eqnarray}
n_{1}^{\left( \psi_{0} \right)}\left( \zeta  \right) &=& {{e}^{2\beta \zeta }}{{\left| {{{\hat{U}}}_{11}}\left( \zeta  \right) \right|}^{2}}\left\langle  \psi_{0} \right|\hat{a}_{1}^{\dagger }\left( 0 \right){{{\hat{a}}}_{1}}\left( 0 \right)\left| \psi_{0} \right\rangle +{{e}^{2\beta \zeta }} \hat{U}_{11}^{*}\left( \zeta  \right){{\hat{U}}_{12}}\left( \zeta  \right)\left\langle  \psi_{0} \right|\hat{a}_{1}^{\dagger }\left( 0 \right){{{\hat{a}}}_{2}}\left( 0 \right)\left| \psi_{0} \right\rangle  \nonumber \\ 
&+&{{e}^{2\beta \zeta }}\hat{U}_{12}^{*}\left( \zeta  \right){{\hat{U}}_{11}}\left( \zeta  \right)\left\langle  \psi_{0} \right|\hat{a}_{2}^{\dagger }\left( 0 \right){{{\hat{a}}}_{1}}\left( 0 \right)\left| \psi_{0} \right\rangle +{{e}^{2\beta \zeta }}{{\left| {{{\hat{U}}}_{12}}\left( \zeta  \right) \right|}^{2}}\left\langle  \psi_{0} \right|\hat{a}_{2}^{\dagger }\left( 0 \right){{{\hat{a}}}_{2}}\left( 0 \right)\left| \psi_{0} \right\rangle \nonumber  \\ 
& +&{{e}^{2\beta \zeta }}\int_{0}^{\zeta }{\int_{0}^{\zeta }{\hat{U}_{11}^{*}\left( \zeta -t' \right){{\hat{U}}_{11}}\left( \zeta -t \right)}}{{e}^{in_{0}^{*}t'}}{{e}^{-i{{n}_{0}}t}}\left\langle \psi_{0} \vert \hat{f}_{1}^{\dagger }\left( t' \right){{{\hat{f}}}_{1}}\left( t \right) \vert \psi_{0} \right\rangle dt'dt  \nonumber \\ 
& +&{{e}^{2\beta \zeta }}\int_{0}^{\zeta }{\int_{0}^{\zeta }{\hat{U}_{12}^{*}\left( \zeta -t' \right){{\hat{U}}_{12}}\left( \zeta -t \right)}}{{e}^{in_{0}^{*}t'}}{{e}^{-i{{n}_{0}}t}}\left\langle \psi_{0} \vert \hat{f}_{2}^{\dagger }\left( t' \right){{{\hat{f}}}_{2}}\left( t \right) \vert \psi_{0} \right\rangle dt'dt, \nonumber \\ \nonumber 
\end{eqnarray}
\begin{eqnarray}
n_{2}^{\left( \psi_{0} \right)}\left( \zeta  \right) &=& {{e}^{2\beta \zeta }}{{\left| {{{\hat{U}}}_{21}}\left( \zeta  \right) \right|}^{2}}\left\langle  \psi_{0} \right|\hat{a}_{1}^{\dagger }\left( 0 \right){{{\hat{a}}}_{1}}\left( 0 \right)\left| \psi_{0} \right\rangle +{{e}^{2\beta \zeta }}\hat{U}_{21}^{*}\left( \zeta  \right){{\hat{U}}_{22}}\left( \zeta  \right)\left\langle  \psi_{0} \right|\hat{a}_{1}^{\dagger }\left( 0 \right){{{\hat{a}}}_{2}}\left( 0 \right)\left| \psi_{0} \right\rangle  \nonumber \\ 
& +&{{e}^{2\beta \zeta }}\hat{U}_{22}^{*}\left( \zeta  \right){{\hat{U}}_{21}}\left( \zeta  \right)\left\langle  \psi_{0} \right|\hat{a}_{2}^{\dagger }\left( 0 \right){{{\hat{a}}}_{1}}\left( 0 \right)\left| \psi_{0} \right\rangle +{{e}^{2\beta \zeta }}{{\left| {{{\hat{U}}}_{22}}\left( \zeta  \right) \right|}^{2}}\left\langle  \psi_{0} \right|\hat{a}_{2}^{\dagger }\left( 0 \right){{{\hat{a}}}_{2}}\left( 0 \right)\left| \psi_{0} \right\rangle  \nonumber  \\ 
&+&{{e}^{2\beta \zeta }}\int_{0}^{\zeta }{\int_{0}^{\zeta }{\hat{U}_{21}^{*}\left( \zeta -t' \right){{\hat{U}}_{21}}\left( \zeta -t \right)}}{{e}^{in_{0}^{*}t'}}{{e}^{-i{{n}_{0}}t}}\left\langle \psi_{0} \vert \hat{f}_{1}^{\dagger }\left( t' \right){{{\hat{f}}}_{1}}\left( t \right) \vert \psi_{0} \right\rangle dt'dt \nonumber  \\ 
&+&{{e}^{2\beta \zeta }}\int_{0}^{\zeta }{\int_{0}^{\zeta }{\hat{U}_{22}^{*}\left( \zeta -t' \right){{\hat{U}}_{22}}\left( \zeta -t \right)}}{{e}^{in_{0}^{*}t'}}{{e}^{-i{{n}_{0}}t}}\left\langle \psi_{0} \vert \hat{f}_{2}^{\dagger }\left( t' \right){{{\hat{f}}}_{2}}\left( t \right) \vert \psi_{0} \right\rangle dt'dt,  \nonumber \\ \label{703_32}
\end{eqnarray}
will have a component related to spontaneous generation, those terms with the integrals, and another to stimulated generation, the rest.
As a practical example, let us use as the initial state a single photon impinging the first waveguide, $\vert \psi_{0} \rangle = \vert 10 \rangle$. Again, we can calculate the mean photon number in the \textbf{balanced gain-loss} dimer, 
\begin{equation}
\begin{array}{c} 
n_{1}^{\left( 10 \right)}\left( \zeta  \right)={{\left| {{{\hat{U}}}_{11}}\left( \zeta  \right) \right|}^{2}} ~-~ 2 \gamma {\int_{0}^{\zeta}{\left| \hat{U}_{11}\left( t \right) \right|}^2} dt, \\
n_{2}^{\left( 10 \right)}\left( \zeta  \right)={{\left| {{{\hat{U}}}_{21}}\left( \zeta  \right) \right|}^{2}} ~-~ 2 \gamma {\int_{0}^{\zeta}{\left| \hat{U}_{21}\left( t \right) \right|}^2} dt,
\end{array}
\end{equation}
and realize that the spontaneous generation terms are identical to those in the spontaneous generation in Equation (\ref{703_14}). 
The same occurs to all the dimer configurations.
In the \textbf{gain-gain} dimer,
\begin{eqnarray}
n_{1}^{\left( 10 \right)}\left( \zeta  \right) &=& {{e}^{2\beta \zeta }}{{\left| {{{\hat{U}}}_{11}}\left( \zeta  \right) \right|}^{2}} ~+~  2 \left( \beta - \gamma \right) {\int_{0}^{\zeta}{\left| \hat{U}_{11}\left( t \right) \right|}^2} e^{2 \beta t} dt \nonumber \\ && + 2 \left( \beta + \gamma \right) {\int_{0}^{\zeta}{\left| \hat{U}_{12}\left( t \right) \right|}^2} e^{2 \beta t} dt,  \nonumber  \\
n_{2}^{\left( 10 \right)}\left( \zeta  \right) &=& {{e}^{2\beta \zeta }}{{\left| {{{\hat{U}}}_{21}}\left( \zeta  \right) \right|}^{2}} ~+~  2 \left( \beta - \gamma \right) {\int_{0}^{\zeta}{\left| \hat{U}_{21}\left( t \right) \right|}^2} e^{2 \beta t} dt \nonumber \\ && +  2 \left( \beta + \gamma \right) {\int_{0}^{\zeta}{\left| \hat{U}_{22}\left( t \right) \right|}^2} e^{2 \beta t} dt, 
\end{eqnarray}
we recover the spontaneous generation terms from Equation (\ref{eq:SGGG}).
In the \textbf{gain-passive} dimer,
\begin{equation}
\begin{array}{c}
n_{1}^{\left( 10 \right)}\left( \zeta  \right) = {{e}^{2\beta \zeta }}{{\left| {{{\hat{U}}}_{11}}\left( \zeta  \right) \right|}^{2}} ~-~  4\gamma {\int_{0}^{\zeta}{\left| \hat{U}_{11}\left( t \right) \right|}^2}  e^{-2 \gamma t} dt,  \\
n_{2}^{\left( 10 \right)}\left( \zeta  \right) = {{e}^{2\beta \zeta }}{{\left| {{{\hat{U}}}_{21}}\left( \zeta  \right) \right|}^{2}} ~-~  4\gamma {\int_{0}^{\zeta}{\left| \hat{U}_{21}\left( t \right) \right|}^2} e^{-2 \gamma t} dt,
\end{array}
\end{equation}
we recover those from the spontaneous generation in Equation (\ref{eq:SGGP}).
Finally, in the \textbf{passive-loss} and \textbf{loss-loss} dimers, the intensity only depends on the initial state and decays due to the nature of the auxiliary $\beta$ parameter,
\begin{equation}
\begin{array}{c}
n_{1}^{\left( 10 \right)}\left( \zeta  \right)={{e}^{2\beta \zeta }}{{\left| {{{\hat{U}}}_{11}}\left( \zeta  \right) \right|}^{2}}, \\
n_{2}^{\left( 10 \right)}\left( \zeta  \right)={{e}^{2\beta \zeta }}{{\left| {{{\hat{U}}}_{21}}\left( \zeta  \right) \right|}^{2}}.
\end{array}
\end{equation}
In these expressions, it is easier to identify that the spontaneous generation component is identical to the one in the vacuum propagation case and the stimulated component,  in the specific case of single photon propagation, will be the same than in the classical dimer, as expected.

\begin{figure}[h]
	\centering
	\includegraphics{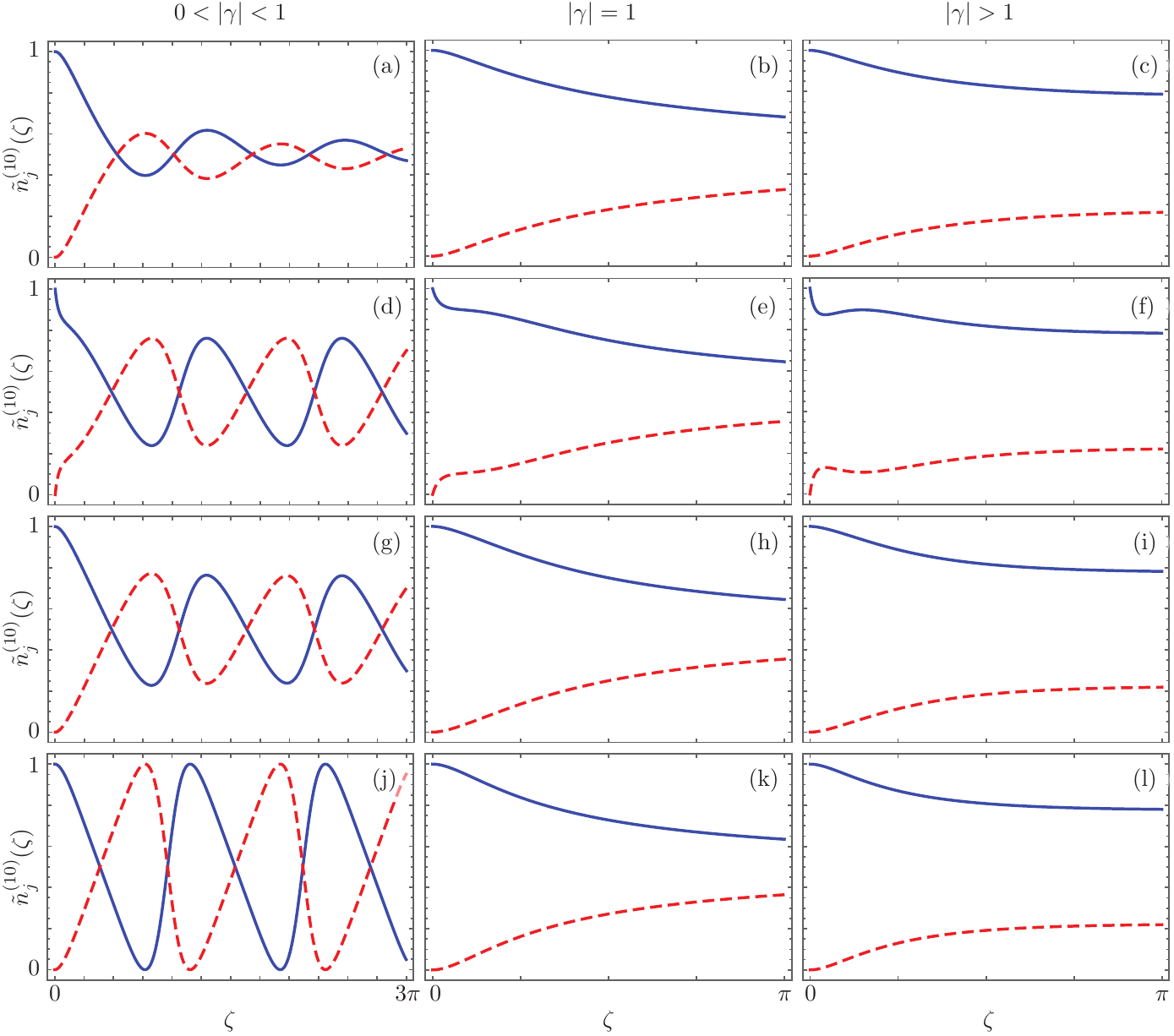}
	\caption{Instantaneously renormalized mean photon number, $\tilde{n}_{j}^{(10)}(\zeta)$, along different realizations of the effective $\mathcal{PT}$-symmetric dimer. The first row, (a)-(c), shows balanced gain-loss, the second row, (d)-(f), shows gain-gain, the third row, (g)-(i), shows gain-passive, and the fourth row, (j)-(l), shows both passive-loss and loss-loss configurations in the $\mathcal{PT}$-symmetric regime, the first column with $\vert \gamma \vert =0.5$, the Kato point with $\vert \gamma \vert = 1$, the second column, and the broken symmetry regime, and the third column with $\vert \gamma \vert =1.2 $. Values for the first and second waveguides are shown with a solid blue and a dashed red lines, in that order.} \label{fig:Fig4}
\end{figure}

Figure \ref{fig:Fig4} shows the renormalized mean photon number for the propagation of a single photon in the first waveguide,
\begin{equation}
\tilde{n}_{j}^{(10)}(\zeta) = \frac{n_{j}^{(10)}(\zeta)}{n_{1}^{(10)}(\zeta)+n_{2}^{(10)}(\zeta)}.
\end{equation}
Note that the initial state is not excluded, as in the spontaneous generation case, because now there will always be a nonzero probability that the dimer will have a photon propagating through it.
Now, as the renormalized mean photon number will have a spontaneous and stimulated component, it is possible to see that the strongest differences in the $\mathcal{PT}$-symmetric regime will occur at small propagation distances, and the oscillation frequency will be larger than in the spontaneous generation case, Fig.~\ref{fig:Fig4}(a), \ref{fig:Fig4}(d), \ref{fig:Fig4}(g). 
In the Kato exceptional point,  Fig. \ref{fig:Fig4}(b), \ref{fig:Fig4}(e), \ref{fig:Fig4}(h), and the broken symmetry regime,  Fig.~\ref{fig:Fig4}(c), \ref{fig:Fig4}(f), \ref{fig:Fig4}(i), the same will happen. 
The strongest deviation from the spontaneous generation signature will occur for small propagation distances and it will take slightly longer propagation distances to reach an asymptotic limit identical in value to that of the spontaneous generation, 
\begin{equation}
\begin{array}{lll}
\lim_{\zeta \rightarrow \infty} \tilde{n}_{1}^{(10)} (\zeta) &=& \frac{1}{2 \gamma} \left( \gamma + \sqrt{ \gamma^2 - 1} \right)^{-1},\\
\lim_{\zeta \rightarrow \infty} \tilde{n}_{2}^{(10)} (\zeta) &=& \frac{1}{2 \gamma} \left( \gamma + \sqrt{ \gamma^2 - 1} \right).
\end{array}
\end{equation}
Finally, it is possible to follow the renormalized mean photon number for the passive-loss and loss-loss dimer which will have identical signatures in all regimes, Fig.~\ref{fig:Fig4}(j), \ref{fig:Fig4}(k), \ref{fig:Fig4}(l), and is able to provide photon localization at any of the waveguides, Fig. \ref{fig:Fig4}(j).

%%%%%%%%%%%%%%%%%%%%%%%%%%%%%%%%%%%%%%%%%%
\section{Photon bunching and anti-bunching in photon propagation}
%%%%%%%%%%%%%%%%%%%%%%%%%%%%%%%%%%%%%%%%%%

We can also study the effect of photon propagation on second order two-point correlations as we did for spontaneous generation. In order to study a different state, we will consider as initial state a $N00N$ state, $\vert N00N \rangle = \left( \vert N0 \rangle + \vert 0N \rangle \right) / \sqrt{2}$, in order to see negative values of the two-point Mandel parameter at least for the initial state,
\begin{eqnarray}
q^{(N00N)}(\zeta) &=&\frac{n^{(N00N)}_{1212}(\zeta)}{n^{(N00N)}_{1}(\zeta) n^{(N00N)}_{2}(\zeta)}-1,
\end{eqnarray} 
where we have defined the following second order two-point correlation function,
\begin{eqnarray}
n^{(N00N)}_{1212}(\zeta) = e^{4 \beta \zeta} \left\langle N00N \vert {{\hat{b}_{1}}^{\dagger }}\left( \zeta \right){{\hat{b}_{2}}^{\dagger }}\left( \zeta \right) \hat{b}_{1}\left( \zeta \right)\hat{b}_{2}\left( \zeta \right) \vert N00N \right\rangle,
\end{eqnarray}
and we take the mean photon numbers as defined in Equation (\ref{eq:MeanPhotonNumber}).

For the two-photon $N00N$ state, $N=2$, the mean photon numbers at the waveguides are provided by the following expressions,
\begin{equation}
\begin{array}{lll}
n^{(2002)}_{1}(\zeta)&=&e^{2 \beta \zeta} {{\left| {{\hat{U}}_{11}}\left( \zeta  \right) \right|}^{2}}+e^{2 \beta \zeta} {{\left| {{\hat{U}}_{12}}\left( \zeta  \right) \right|}^{2}}+n^{(00)}_{1}(\zeta), \\
n^{(2002)}_{2}(\zeta)&=&e^{2 \beta \zeta} {{\left| {{\hat{U}}_{21}}\left( \zeta  \right) \right|}^{2}}+e^{2 \beta \zeta} {{\left| {{\hat{U}}_{22}}\left( \zeta  \right) \right|}^{2}}+n^{(00)}_{2}(\zeta),
\end{array}
\end{equation}
where the first and second terms corresponds to the stimulated generation, and the third term is related to the spontaneous generation and can be recovered from Section \ref{sec:SpontaneousGeneration} for each and every configuration.
The general expression for the two-point correlation is summarized in the following,
\begin{eqnarray}
n_{1212}^{(2002)}(\zeta)&=&e^{4 \beta \zeta }{{\left| {{\hat{U}}_{11}}{{\hat{U}}_{21}}+{{\hat{U}}_{12}}{{\hat{U}}_{22}} \right|}^{2}}  +  {n^{(00)}_{1}(\zeta)}{n^{(00)}_{2}(\zeta)}+ \vert n^{(00)}_{12}(\zeta) \vert^{2} \nonumber \\ 
& & + e^{2 \beta \zeta } \left[ {n^{(00)}_{1}(\zeta)}\left( {{\left| {{\hat{U}}_{21}} \right|}^{2}}+{{\left| {{\hat{U}}_{22}} \right|}^{2}} \right)+ {n^{(00)}_{2}(\zeta)}\left( {{\left| {{\hat{U}}_{11}} \right|}^{2}}+{{\left| {{\hat{U}}_{12}} \right|}^{2}} \right) \right] \nonumber \\ 
& & + 2~e^{2 \beta \zeta }~\Re\left[ {n^{(00)}_{12}(\zeta)}\left( \hat{U}_{21}^{*}{{\hat{U}}_{11}}+\hat{U}_{22}^{*}{{\hat{U}}_{12}} \right)\right].
\end{eqnarray}
For the \textbf{balanced gain-loss} dimer, we use the spontaneous generation terms, $n^{(00)}_{j}(\zeta)$, provided by Equation (\ref{703_14}), and the first order two-point correlation, $n^{(00)}_{12}(\zeta)$, from Equation (\ref{703_23}). In the \textbf{gain-gain} dimer, the expressions for $n^{(00)}_{j}(\zeta)$ and $n^{(00)}_{12}(\zeta)$ are given by Equations (\ref{eq:SGGG}) and  (\ref{eq:1003_24}), in that order. In the \textbf{gain-passive} dimer, we only need the definitions provided by Equations (\ref{eq:SGGP}) and (\ref{eq:12GP}) for the spontaneous generation and the first order two-point correlation, respectively. 
Finally, for \textbf{passive-loss} and \textbf{loss-loss} dimers, the spontaneous generation is null, $n^{(00)}_{j}(\zeta)=n^{(00)}_{12}(\zeta)=0$, such that 
\begin{eqnarray}
n_{1212}^{(2002)}(\zeta) = e^{4 \beta \zeta }{{\left| {{\hat{U}}_{11}}{{\hat{U}}_{21}}+{{\hat{U}}_{12}}{{\hat{U}}_{22}} \right|}^{2}}.
\end{eqnarray}
Obviously, the adequate parameter $\beta$ from Table~\ref{tab:Tab1} should be used for each configuration.
These expressions become complicated enough that we must rely on a figure-based analysis.

\begin{figure}[h]
	\centering
	\includegraphics{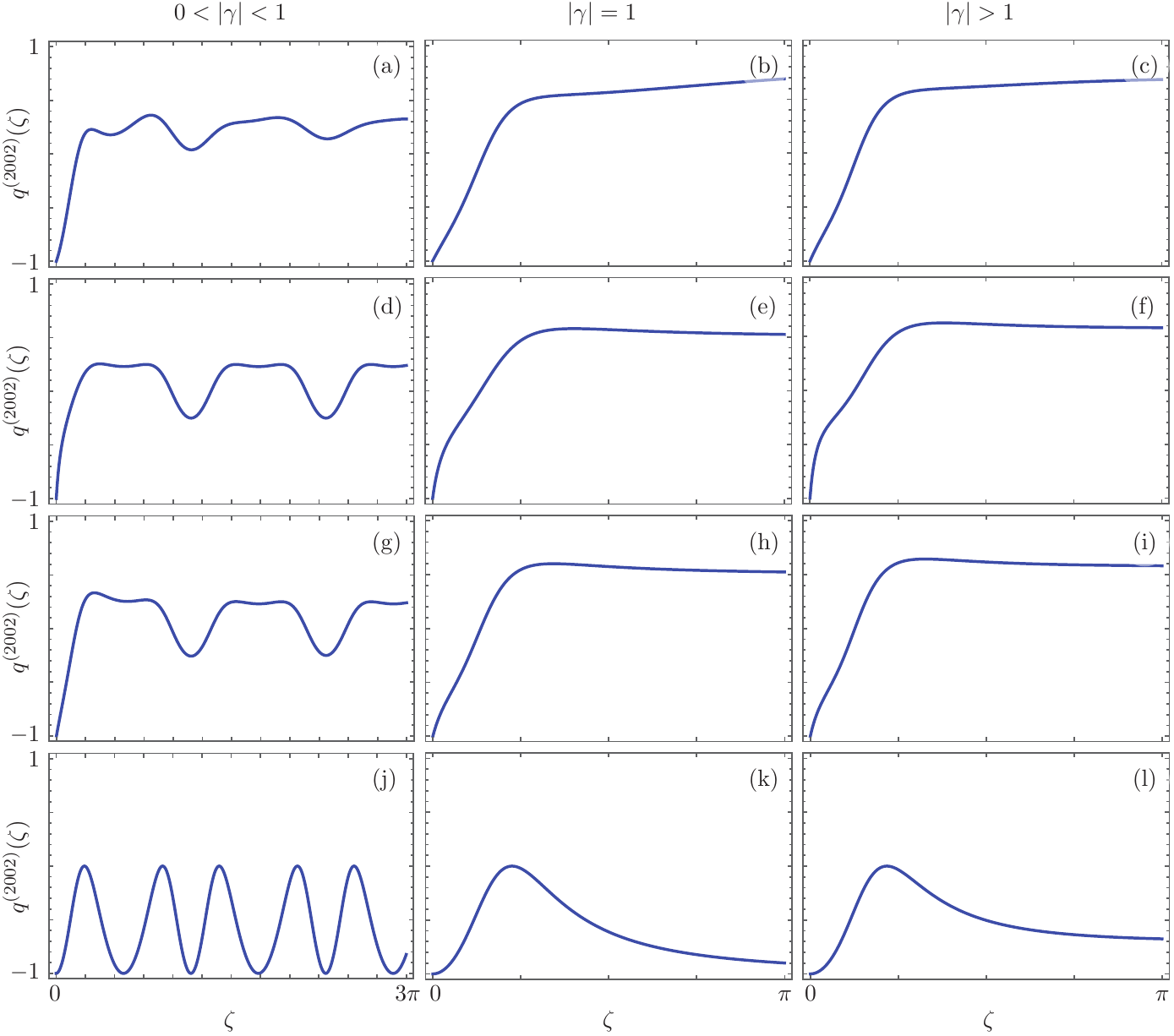}
	\caption{Photon bunching and anti-bunching shown in terms of the $q^{(2002)}(\zeta)$ parameter for different realizations of the effective $\mathcal{PT}$-symmetric dimer. The first row, (a)-(c), shows balanced gain-loss, the second row, (d)-(f), shows gain-gain, the third row, (g)-(i), shows gain-passive, and the fourth row, (j)-(l), shows both passive-loss and loss-loss configurations in the $\mathcal{PT}$-symmetric regime, the first column with $\vert \gamma \vert =0.5$, the Kato point with $\vert \gamma \vert = 1$, the second column, and the broken symmetry regime, the third column with $\vert \gamma \vert =1.2 $.} %there is no explanation for subfigure
	\label{fig:Fig5}
\end{figure}

Figure \ref{fig:Fig5} shows the two-point Mandel parameter for the different dimer configurations in the $\mathcal{PT}$-symmetric, the Kato exceptional point, and broken $\mathcal{PT}$-symmetry regimes. 
Now, the initial state shows anti-bunching, a negative value of the two-point Mandel parameter, due to its delocalization of the two-photon state. 
For the balanced gain-loss, the initial state propagates and, after a critical propagation distance, presents bunching in the $\mathcal{PT}$-symmetric, Fig.~\ref{fig:Fig5}(a), the Kato exceptional point, Fig. \ref{fig:Fig5}(b), broken symmetry, Fig. \ref{fig:Fig5}(c), and regimes.
The gain-gain and gain-passive dimers show a similar, more interesting behavior where the propagated state oscillates between an anti-bunched and bunched state in the $\mathcal{PT}$-symmetric regime, Fig. \ref{fig:Fig5}(d), \ref{fig:Fig5}(g), and the transition from an anti-bunched to a bunched state in the Kato exceptional point, Fig. \ref{fig:Fig5}(e), \ref{fig:Fig5}(h), and broken symmetry regimes, Fig.~\ref{fig:Fig5}(f),~\ref{fig:Fig5}(i).
Finally, in both the passive-loss and loss-loss dimers, the probability of losing photons 
%It was ''lossing'' - Pls confirm
makes the propagated state anti-bunched in all regimes, Fig. \ref{fig:Fig5}(j)--\ref{fig:Fig5}(l). 

%%%%%%%%%%%%%%%%%%%%%%%%%%%%%%%%%%%%%%%%%%
\section{Conclusions}
%%%%%%%%%%%%%%%%%%%%%%%%%%%%%%%%%%%%%%%%%%

We have calculated the propagation of photon states through non-Hermitian linear dimers with Gaussian gain and losses. 
As a practical example, we studied propagation in the different experimentally feasible configurations of the $\mathcal{PT}$-symmetric dimer and show that each and every configuration presents a different signature in the propagation of vacuum, single and two-photon states. 
These signatures go beyond the mean photon number at the waveguides, which can be split into spontaneous and stimulated generation components, and can be found, for example, in the second order two-point correlation of the photon state propagating through the dimers. 

First, let us focus on the mean photon number signatures. They show that the propagation of vacuum and single photons through dimers in the $\mathcal{PT}$-symmetry regime might provide us with directional coupler devices controllable by the propagation length. Furthermore, devices that provide full single photon switching can only be designed using dimers in the passive-loss and loss-loss configuration. Outside the $\mathcal{PT}$-symmetric regime, the asymptotic stability of the dimers in any given configuration suggest their use as symmetric intensity sources at the Kato exceptional point and, in the broken symmetry regime, they might provide asymmetric intensity sources where the intensities ratio can be controlled by the linear properties of the material.  

On the other hand, second order two-point correlation signatures can help us choose configurations depending on the type of state needed for a particular application. For example, if spatially separable states are needed, choosing a configuration showing photon bunching comes naturally.  Spatially entangled states are provided by passive-loss and loss-loss configurations where the initial two-photon state is delocalized in the waveguides.

We want to note that our description of linear media is far from complete. Real world linear materials saturate; this induces further restrictions on the model that we do not consider in this manuscript and could provide further avenues of research. Furthermore, the addition of new interactions, like introducing a two-level system to create more complex hybrid devices \cite{Lepert2011p113002}, requires a first principle reformulation in order to recover adequate effective models.

%%%%%%%%%%%%%%%%%%%%%%%%%%%%%%%%%%%%%%%%%%
\vspace{6pt} 

%%%%%%%%%%%%%%%%%%%%%%%%%%%%%%%%%%%%%%%%%%
%% optional
%	\supplementary{The following are available online at www.mdpi.com/link, Figure S1: title, Table S1: title, Video S1: title.}

%%%%%%%%%%%%%%%%%%%%%%%%%%%%%%%%%%%%%%%%%%
\ack{Jos\'e Delfino Huerta Morales acknowledges financial support from CONACYT $\#$294921 PhD grant.}

\section*{References}
\bibliographystyle{unsrt}
%\bibliography{references}

\end{document}